# Nanograin Density Outside Saturn's A ring

Robert E. Johnson[1,2], Wei-Lin Tseng[3], M.K. Elrod[4,5], A.M. Persoon[6]

[1] Engineering Physics, University of Virginia, Charlottesville, VA 22902; 434-422-2424; rej@virginia.edu; http://orcid.org/0000-0001-7798-5918
[2] Physics, NYU, NY, NY 10003
[3] National Taiwan Normal University; No. 88, Sec. 4, Tingzhou Rd., Wenshan District, Taipei 11677, Taiwan (R.O.C.): 886-2-77346402; wltseng@ntnu.edu.tw
[4] NASA Goddard Space Flight Center, Greenbelt, MD, 20771. 757-725-3014 meredith.k.elrod@nasa.gov
[5] CRESST, University of Maryland, College Park, MD 20742
[6] Dept. of Physics and Astronomy, University of Iowa, Iowa City, IA, 52242 ann-persoon@uiowa.edu

**Abstract**

The observed disparity between the radial dependence of the ion and electron densities measured by the Cassini plasma (CAPS) and radio (RPWS) science instruments are used to show that the region between the outer edge of Saturn's main rings and its tenuous G-ring is permeated with small charged grains (nanograins). These grains emanate from the edge of the A ring and from the tenuous F and G rings. This is a region of Saturn's magnetosphere that is relatively unexplored, but will be a focus of Cassini's F ring orbits prior to the end of mission in September 2017. Confirmation of the grain densities predicted here will enhance our ability to describe the formation and destruction of material in this important region of Saturn's magnetosphere.

**Introduction**

As the enormously successful Cassini mission winds down, the focus will be on two regions in the Saturnian system that have mostly been avoided due to possible hazards. From November 2016 through April 2017 Cassini will carry out ~20 orbits outside the edge of Saturn's main rings (~2.1$R_S$; 1$R_S$ = Saturn's radius) and close to the tenuous F ring (~2.23 $R_S$) focusing on the environment between the main rings and the tenuous rings. Subsequent to these orbits, referred to as the F ring orbits, Cassini will then change to orbits that are inside the D ring (< 1.1$R_S$) in order to study Saturn's extended atmosphere and its interaction with the rings, the so-called proximal orbits. Here we show that, in addition to the presence of grains observed in the tenuous F and G rings, plasma measurements indicate that there is a significant density of small charged grains that permeate the region outside the visible edge of Saturn's A ring.

On Saturn Orbit Insertion (SOI) in 2004, Cassini's plasma instrument discovered a surprisingly robust density of ionized oxygen molecules over the rings (Young et al; 2005; Tokar et al. 2005) produced by photoionization of an oxygen ring atmosphere. Because of the low radiation flux over the rings (e.g., Cooper et al. 2016), $O_2$ is formed primarily by solar UV-induced decomposition of the icy ring particles with a concomitant, but much more extended, $H_2$ atmosphere (Johnson et al. 2006). These molecules are scattered from over the rings into Saturn's magnetosphere and into its atmosphere by ion-molecule interactions (Johnson et al. 2006; Luhmann et al. 2006; Bouhram et al. 2006; Martens et al. 2008).

Cassini subsequently crossed the ring plane outside of the A ring where it observed a much higher density thermal plasma. In spite of the high background radiation (Tokar et al. 2005), plasma ion densities and composition were extracted (Elrod et al. 2010; 2014) indicating that the newly discovered ring atmosphere extended well beyond the outer edge of the A ring. Although Cassini never came closer than ~2.4$R_S$, the analysis by Elrod et al. for orbits with a periapsis inside of that of Mimas (~3$R_S$) confirmed that the ring atmosphere was primarily produced by the solar UV decomposition of ice (Johnson et al. 2006). That is, because the tilt of the ring plane to the solar flux varies during Saturn's orbit, an atmosphere formed by solar UV should exhibit a seasonal dependence (Tseng et al. 2010). The variation in the CAPS thermal plasma $O_2^+$ data from near southern summer solstice to equinox clearly exhibited such a dependence (Elrod et al. 2012; 2014). A seasonal dependence was also seen in the energetic particle data (Christon et al 2013; 2014) and the plasma electron data (Persoon et al. 2015). Although there is now strong evidence for a seasonal varying ring atmosphere, the region between the outer edge of the A ring and inside the F ring, through which Cassini will pass during its F ring orbits, is not well described (e.g., Cooper et al. 2016). We show here that, in addition to the ionization of molecules scattered from over the rings and the ionization of neutrals diffusing inward from Enceladus, this region is permeated by charged nanograins likely coming from the edge of main rings and the tenuous rings.

**Cassini Plasma Data**

The near equatorial CAPS ion densities (Elrod et al. 2012; 2014) and the corresponding electron densities extracted from the RPWS data (Persoon et al. 2015) for the Cassini SOI orbit are displayed in Fig. 1a as a function of radial distance, $R$, from Saturn. These both exhibit striking spatial morphologies, but the densities disagree by up to an order of magnitude unlike at larger radial distances where the RPWS and CAPS charge densities are reasonably consistent (e.g., Persoon et al. 2015; Elrod et al. 2012; Sittler et al. 2006). In Fig. 1b we also show results from a later orbit during which Cassini penetrated into this region. The large drop in the ion density from SOI in 2004 to that in 2007 was attributed the change in the orientation of the ring plane to the solar flux as Saturn approaches equinox, the seasonal variation discussed above. More important to the work here, an unpredicted, steep decrease in the ion density with decreasing distance from the edge of the rings is seen in both Fig.1a and 1b. It is also seen in Fig. 1b, that at the smaller $R$ the free electron density dropped below the limit needed to stimulate the RPWS upper hybrid resonance. These are points we will return to below.

The physical processes occurring in this region were examined in Tseng et al. (2013). Here we revisit that work focusing first on the use of the near equatorial plasma data to predict the presence of a significant density of as yet unobserved small charged grains in the region between the outer edge of the A ring and the G ring and then on the role of these grains in quenching the ions. Although the effect of grains on plasma transport in this region has been discussed (e.g., Sakai et al. 2013; Roussos et al. 2016), the data in Fig. 1 was not considered.

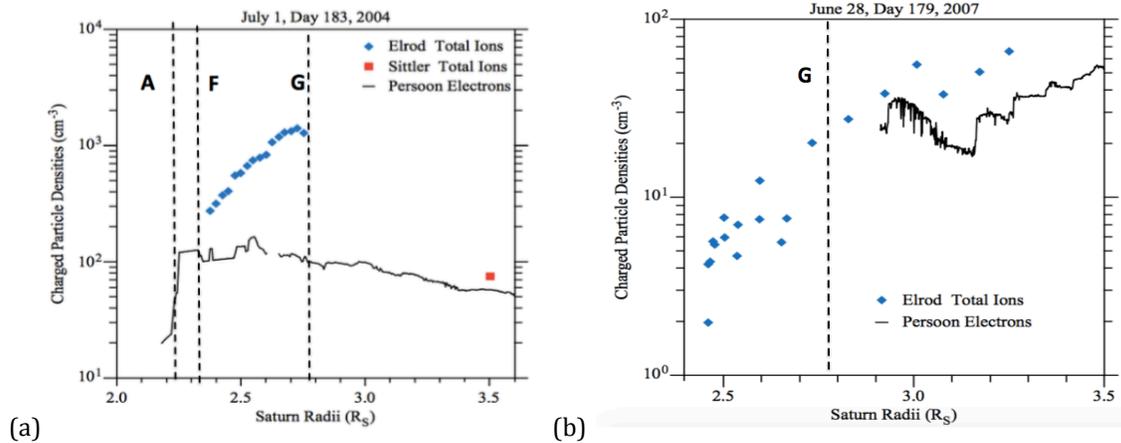

(a) (b)

Fig. 1 Ion and electron densities within ~ 0.5$R_S$ of the equator (a) SOI orbit; (b) a 2007 orbit. Lines: electron density extracted from the RPWS upper hybrid resonance (Persoon et al. 2015); blue diamonds: total ions from CAPS: $O_2^+$ + $W^+$ dominated by $O_2^+$ at SOI and $W^+$ at 2007 (Elrod et al. 2014); red square: total $H_2^+$ + $W^+$ from CAPS dominated by $W^+$ (Sittler et al. 2006)

**Nanograin Grain Density**

The region between the G ring and the outer edge of the A ring is primarily populated by two neutral gas sources, the gas torus produced by the Enceladus plumes and the extended ring atmosphere, with contributions from the ice particles and small moons in the F and G rings. However, combining models of these two principal sources with the ionization rates and models for diffusion cannot explain the plasma observations in Fig. 1. Although for most regions of Saturn's magnetosphere the ion and electron densities extracted show reasonable agreement, this is not the case, for instance, within the Enceladus plumes where the disparity was attributed to the presence of small charged ice grains (Morooka et al 2011; Hill et al. 2012). Here we propose that the disagreement in Fig. 1a also indicates the presence of a significant density of charged grains with sizes well below the limit detected by the dust instrument on Cassini.

In this region the electron temperature is low, so that grains are predominantly negatively charged (Jurac et al. 1995; Kempf et al. 2006). The RPWS upper hybrid resonance is sensitive only to the free electrons. Unfortunately, the electron instrument on the CAPS suite, which detected negatively charged clusters at Titan (Coates et al. 2010) and small charged grains near Enceladus (Hill et al. 2012) did not obtain electron densities due to spacecraft charging and the significant background radiation (e.g., Schippers et al. 2009). Because of this background, only the heavy ion densities were extracted from the CAPS measurements, the $O_2^+$ and the water group ions, $W^+$ ($H_3O^+$, $H_2O^+$, $OH^+$, $O^+$). Rough upper limits were placed on the light ions ($H_2^+$, $H^+$) indicating they were a small fraction of the near equatorial positive charge (Elrod et al. 2014) consistent with models (Tseng et al. 2011).

The ion density in Fig. 1a is dominated by $O_2^+$ with temperatures close to the pick-up energy, indicating the ions are short-lived and freshly formed, primarily by ionization of $O_2$ in the extended ring atmosphere. The ion densities in the 2007 data in Fig. 1b are dominated by $W^+$ with a temperature that is a fraction of the pick-up energy. This plasma is formed primarily by ionization of water molecules in the Enceladus gas torus with a significant but smaller contribution from the $O_2$ ring atmosphere. It is also seen in Fig. 1 that there is a sharp drop in the heavy ion charge density with decreasing distance from the edge of the A-ring. This radial dependence is *very* different from that predicted by models of the ion source strength and radial diffusion, indicative of a strong quenching process (Tseng et al. 2013). That is, accounting for both the ring and Enceladus sources the ion production rate varies slowly with *R* across this region, so that the loss processes and not the source processes determine the observed radial dependence. Since the difference between the ion and free electron densities, as well as the rapid quenching, can be due to the presence of negatively charged grains, we estimate their density and size distribution in anticipation of the F ring orbits.

The ion density goes through a maximum as suggested by the peak in Fig. 1a and by the much lower density at ~3.8Rs where the CAPS measurements are roughly comparable to the RPWS electron densities (e.g., Sittler et al. 2006). Such a maximum is consistent with the ionization of a robust $O_2$ atmosphere emanating from the edge of the A-ring, but rapidly quenched close to the ring source. This is similar to the observation that the plasma density produced by the Enceladus gas torus peaks outside the neutral source region (Sittler et al. 2006; Tokar et al. 2008). Below, we first estimate the grain density consistent with the observation in Fig. 1a and then show they act to quench the ions.

Assuming the difference between the ion and electron densities, $\Delta n_q$, is due to distribution of small negatively charged grains, then $\Delta n_q$ depends on the density of grains, $n_g$, and the charge per grain. The average charge per grain in turn depends on the grain radius, $r_g$, and the plasma environment. Based on the plasma properties, models suggest the grain potential is ~ -1 to -2eV in this region (Jurac et al. 1995; Kempf et al. 2006). As these very small grains are likely singly charged (Hill et al. 2012), setting $n_g \sim \Delta n_q$ gives a rough upper bound to the small negatively charged grain density. At SOI, $\Delta n_q$ peaks at ~ 1000/cm$^3$ just inside the G ring orbit.

Grain size distributions are predicted to be steeply varying, so that the number density of grains with radius greater than *a* is estimated as $n_g \sim n_{g0} (r_g/a)^{-\alpha}$ with $\alpha \sim 4-5$ and $n_{g0}$ a normalization density (Kempf et al., 2008). The peak in $\Delta n_q$ occurs near 3R$_S$ where the Cassini Dust Analyzer (CDA) gives a density for $r_g > \sim$ 1$\mu m$ grains of $n_g \sim 3 \times 10^{-9}$/cm$^3$. Using this to normalize $n_g$, we find that a density, $\Delta n_q$, of singly charged grains with a minimum grain radius, $r_{gmin} \sim 1\mu m [n_{g0}/\Delta n_q]^{1/\alpha}$ can account for the difference in the observed charge densities. For $\Delta n_q \sim 1000$cm$^{-3}$ at 2.7R$_S$, $r_{gmin} \sim 1.3$*nm* for $\alpha = 4$ and ~5*nm* for $\alpha = 5$. Since the distribution is strongly peaked, the dominant grain size is within a factor of a few times these estimates for $r_{gmin}$. It also means that these estimates do not change enormously if the larger grains are more highly charged, as one would expect to be the case. Therefore, the difference near the peak of the CAPS data for SOI can be accounted for by singly charged nanograins with a size distribution like that discussed above. Nearer to the F ring, $\Delta n_q \sim 100$ cm$^{-3}$. In the absence of CDA measurements in this region, we use the same normalization for the grain size distribution, in which case $r_{gmin}$ again does not change significantly due to the steep size distribution. These densities are much larger than the E ring grain densities in this region and much larger than the grain densities extrapolated by Sakai et al (2013) to the edge of the A ring (~0.01-0.03 cm$^{-3}$) in order to explain the drag on the measured ion flow speeds.

The above estimates indicate that the ion and free electron differences in Fig. 1a could be accounted for by a density of very small, as yet, unobserved nanograins. However, this conclusion must be consistent with the observation of the steep radial gradient seen in the ion data. Using an electron temperature of ~ -2eV, Tseng et al. (2013) estimated that loss due to electron-ion recombination gave ion lifetimes > ~10$^5$s assuming a free electron density equal to the ion density. Using the lower free electron densities in Fig 1a, lengthens this time. However, ions can also collide with and neutralize on grains. Because the grain potential is much smaller that the ion energies, and ignoring the attractive force between the positive ion and negatively charged grains, the ion-grain collision cross section can be estimated from the grain radius: $\sigma_{col} \sim \pi r_g^2$. As the measured ion flow speeds at SOI are of the order of the co-rotation speed (Elrod et al. 2012) we use an average, $<v_{col}> \sim 10$km/s, as a rough estimate of the difference between the ion co-rotation and the Kepler speeds in this region. The collision rate is then obtained by integrating over the grain size distribution:

$$\int v_{col}\, \sigma_{col}\, (dn_g/dr_g)\, dr_g \sim \pi r_{gmin}^2 [\alpha/(\alpha-2)]\, \Delta n_q <v_{col}> \qquad (1)$$

Using the SOI values determined above, the collision time is ~10$^4$s for $\alpha$ = 4 and ~800s for $\alpha$ = 5. Such times are *much* shorter than those estimated for recombination, as discussed above, as well as for the other loss processes in Tseng et al. (2013).

Since the neutral density and, hence, the O$_2^+$ source rate decrease slowly with increasing *R* outside the A ring, we propose the steep decrease with decreasing *R* seen in Fig.1a in the SOI ion density, is due to the quenching of the ions by collisions with nanograins being emitted from the edges of the A rings and the tenuous rings. Consistent with this, simulations indicate that small charged grains generated in this region have relatively long lifetimes (e.g., Liu and Ip 2014). The much shorter plasma ion lifetime than that estimated earlier suggests that the O$_2$ ring atmosphere is more robust (Johnson et al. 2006) or that the ionization in the region is dominated by presence of a small, but hot, electron population (Tseng et al. 2013). After correcting the CAPS electron data for the significant background, the electron energy spectra did exhibit a hot component outside the edge of the A ring (Fig. 1b in Schippers et al. 2009), although absolute values were not extracted,

Unfortunately for the 2007 orbit, the results in Fig.1b cannot be used to determine the difference between the ion and the electron densities close to the F ring. The loss in the signal in the upper hybrid observations near ~2.9R$_S$ indicates the free electron density dropped below ~4/cm$^3$ outside the G ring. Because the electron-cyclotron frequency increases with increasing field strength as *R* decreases, the lack of an upper hybrid detection implies a decreasing upper limit on the free electron density with decreasing $R$. Therefore, the free electron density remained well below the ion density throughout the region inside about 2.9$R_S$ consistent with the presence of small charged grains.

## Summary

That there are micron size or larger grains in this region is, of course, well-known from the CDA measurements and by light scattering in the tenuous rings. But the difference between the electron and ion densities, and the steep drop in the ion density with decreasing $R$, is very different from the radial dependence of the ion source term. Here we show that these plasma measurements can be explained by the presence of negatively charged nanograins that reduce the free electron density and act to collisionally quench the ions. Both the SOI and 2007 data indicate this region is permeated with such grains and their density increases with decreasing $R$. These grains are likely sourced from the edge of the A ring as well as from the F and G rings. This is consistent with studies indicating that collisions of small icy bodies inside 3.0Rs produce debris (Tiscareno et al., 2013; Attree et al., 2014) and it is even the case that small objects can form from such debris, as seen at the edge of the A ring (Murray et al., 2014).

Orbiting debris is known to deplete the energetic ion and electron fluxes (e.g., Cuzzi and Burns, 1988; Paranicas et al., 2008; Cooper et al. 2016). More recently Roussos et al., (2016) suggested that energetic particle signatures inside the orbit of Mimas appear to be shifted by plasma transport affected by the presence of charged grains. Here we have used the Cassini CAPS and RPWS data to add to that picture by estimating the charged nanograin density in this region. Since the F ring orbits will occur as Saturn approaches the northern hemisphere summer solstice, we expect plasma densities of, roughly, the same order of magnitude as those estimated from the SOI data. Unfortunately, the CAPS instrument will be off. Therefore, ion densities must be derived from the Ion Neutral Mass Spectrometer (INMS), but, since the electron densities are expected to be much larger than those in the 2007 data, the upper hybrid resonance should be observable. Grain densities might be detectable by the INMS, by energetic particle absorptions or, possibly, by plasma waves generated by impacts on the spacecraft (e.g., Schippers et al. 2014). The results presented here give, for the first time, an estimate of the nanograin density in the vicinity of the F ring, which can be used, along with the expected Cassini measurements, to constrain modeling of the formation and destruction of material in this important region of Saturn's magnetosphere.

## Acknowledgements

REJ acknowledges support from NASA via the Cassini mission through SwRI and JPL. The Cassini RPWS data are in the Planetary Data System. W.-L.T. acknowledges support from MOST 104-2112-M-003 -006 -MY2 in Taiwan (R.O.C.). AMP acknowledges support by NASA through contract 1415150 with the Jet Propulsion Laboratory